\documentclass{iaus}
\usepackage{graphicx}
\def\rd{$r_d$}

\title[Scale Length of Disk Galaxies] 
{Scale Length of Disk Galaxies}

\author[Kambiz Fathi]   
{Kambiz Fathi$^1$}

\affiliation{$^1$Stockholm Observatory, Department of Astronomy, Stockholm University, Sweden\\email: {\tt kambiz@astro.su.se}}

\pubyear{2011}
\volume{277}  
\pagerange{\ }
\jname{Tracing the Ancestry of Galaxies on the Land of our Ancestors}
\editors{Editors: C. Carignan, F. Combes, K. C. Freeman}
\begin{document}

\maketitle

\begin{abstract}
Disk scale length \rd\ and central surface brightness $\mu_0$ for a sample of 29955 bright disk galaxies from the Sloan Digital Sky Survey have been analysed. Cross correlation of the SDSS sample with the LEDA catalogue allowed us to investigate the variation of the scale lengths for different types of disk/spiral galaxies and present distributions and typical trends of scale lengths all the SDSS bands with linear relations that indicate the relation that connect scale lengths in one passband to another. We use the volume corrected results in the $r$-band and revisit the relation between these parameters and the galaxy morphology, and find the average values $\langle r_d\rangle = 3.8\pm 2.1$ kpc and  $\langle\mu_0\rangle=20.2\pm 0.7$ mag~arcsec$^{-2}$. The derived scale lengths presented here are representative for a typical galaxy mass of $10^{10.8} \rm{~M}_\odot$, and the RMS dispersion is larger for more massive galaxies. We analyse the \rd--$\mu_0$ plane and further investigate the Freeman Law and confirm that it indeed defines an upper limit for $\mu_0$ in bright disks ($r_\mathrm{mag}<17.0$), and that disks in late type spirals ($T \ge 6$) have fainter central surface brightness. Our results are based on a sample of galaxies in the local universe ($z< 0.3$) that is two orders of magnitudes larger than any sample previously studied, and deliver statistically significant results that provide a comprehensive test bed for future theoretical studies and numerical simulations of galaxy formation and evolution.
\end{abstract}

\firstsection 
\section{Overview}
The mass distribution of a disk is set by the \rd\ and in the exponential case, 60\% of the total mass is confined within two scale lengths and 90\% within four scale lengths. Moreover, the angular momentum of a disk is set by \rd\ and the mass distribution of its host halo, and the fact that the angular momentum vectors are aligned suggests that there is a physical relation between the two. During the formation process, mergers and associated star formation and feedback processes play a crucial role in the resulting structure, however, the observed sizes of disks suggest that the combination of these physical processes yield that galactic disks have not lost much of the original angular momentum acquired from cosmological torques (White \& Rees 1978). A large \rd\ disk forms when the disk mass is smaller than the halo mass over the disk region, and vice versa, a small \rd\ disk forms when the mass of the disk dominates the mass of the halo in any part of the disk. The self gravitating disk will also modify the shape of the rotation curve near the centre of a galaxy and the disk is then set to undergo secular evolution. The natural implication of this scenario is that the \rd\ dictates the life of a disk, and consequently, is a prime factor which determines the position of a galaxy on the Hubble sequence. 

Here we analyse the \rd\ and $\mu_0$ from an unprecedentedly large sample of bright disk galaxies in the nearby universe ($z< 0.3$) using the Sloan Digital Sky Survey (SDSS) Data Release 6 (York et al. 2000; Adelman-McCarthy et al. 2008). We have used the Virtual Observatory tools and services to retrieve data in all ($u$, $g$, $r$, $i$, and $z$) SDSS band and used the LEDA catalogue (Paturel et al. 2003) to retrieve morphological classification information about our sample galaxies, and those with types defined as Sa or later are hereafter refereed to as disk galaxies. In the $g$, $i$, and $z$-band, $\approx 27000$--30000 galaxies were analysed, and in the $u$-band, \rd\ and $\mu_0$ were robustly derived for a few hundred objects. Throughout this presentation, we use disk parameters in the $r$-band to provide a comprehensive test bed for forthcoming cosmological simulations (or analytic/semi-analytic models) of galaxy formation and evolution. Further details have been presented in Fathi et al. (2010) and Fathi (2010).

One prominent indicator for a smooth transition from spiral toward S0 and disky ellipticals is provided by the \rd--$\mu_0$ diagram where $\mu_0$ is the central surface brightness of the disk, where spirals and S0s are mixed and disky ellipticals populate the upper left corner of this diagram. Another instructive relation is the Freeman Law (Freeman 1970) which relates $\mu_0$ to the galaxy morphological type. Although, some studies have found that the Freeman Law is an artefact due to selection effects (e.g., Disney et al. 1976), recent work have shown that proper consideration of selection effects can be combined with kinematic studies to explore an evolutionary sequence. In the comparison between theory and observations, two issues complicate matters. On the theory side mapping between initial halo angular momentum and \rd\ is not trivial, partly due to the fact that commonly the initial specific angular momentum distribution of the visible and dark component favour disks which are more centrally concentrated disks than exponential. Observationally, comprehensive samples have yet not been studied, and the mixture of different species such as low and high surface brightness galaxies complicate the measurements of disk parameters.

\begin{figure*}
\begin{center}
 \includegraphics[width=.69\textwidth]{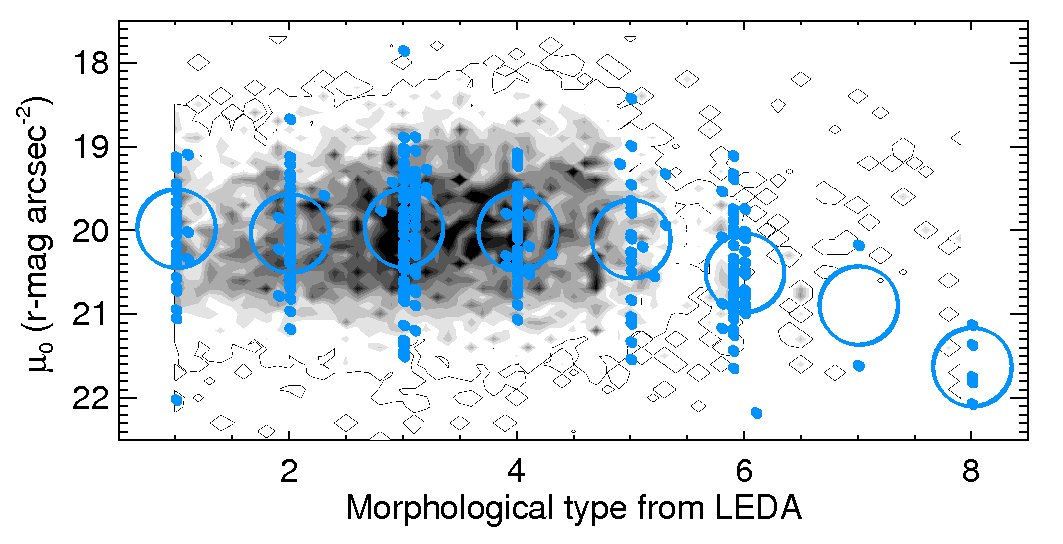}
  \caption{Volume corrected distribution of disk central surface brightness $\mu_0$ versus morphological type T. The blue dots illustrate the 282 galaxies for which robust morphological types were given in the LEDA catalogue (type error smaller than 0.5) and the open circles show the average $\mu_0$ for each type.} 
\label{fig:freeman}
\end{center}
\end{figure*}
\begin{figure*}
\begin{center}
 \includegraphics[width=.99\textwidth]{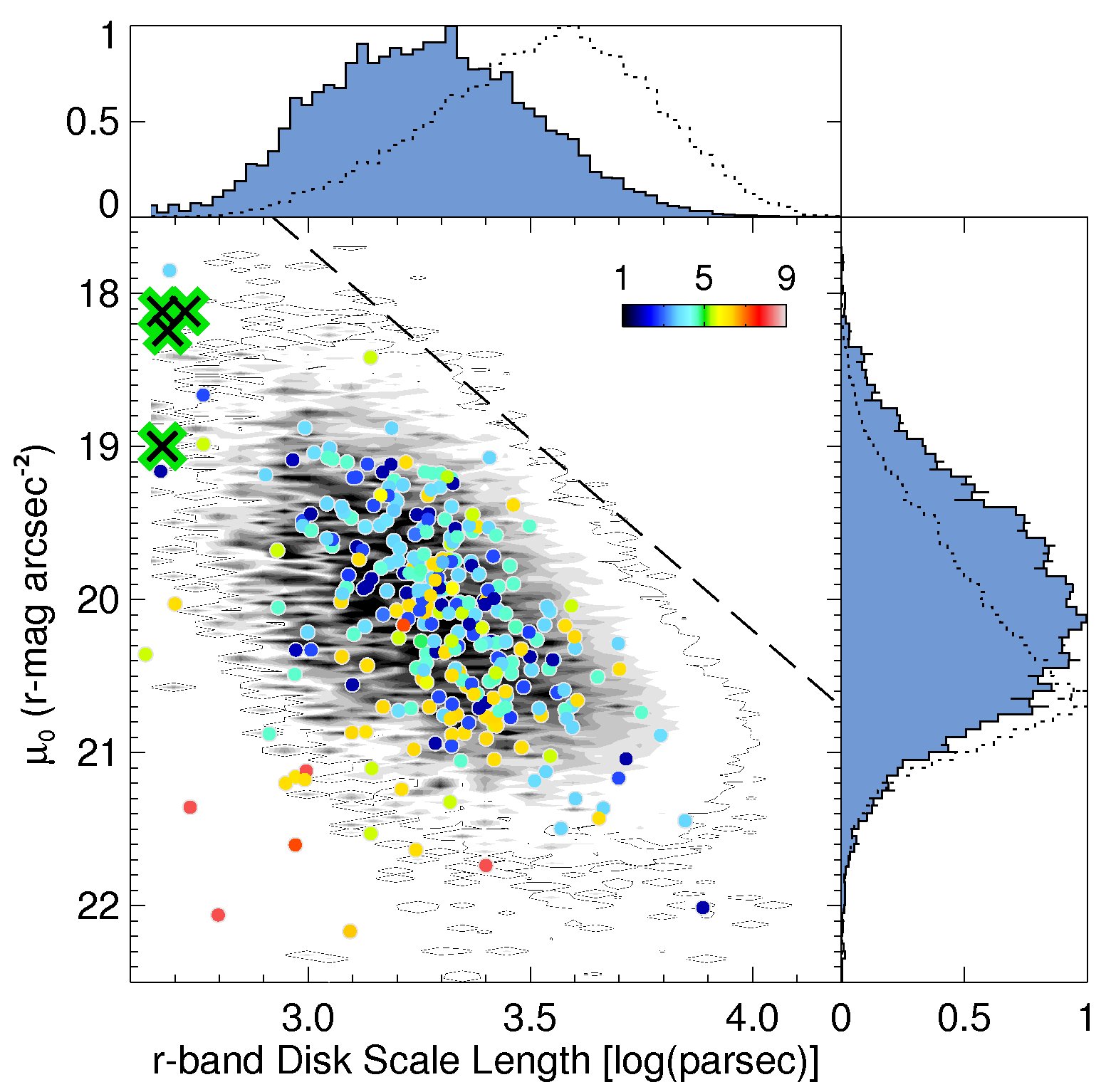}
  \caption{The \rd--$\mu_0$ plane of our sample, with the 282 galaxies with reliable morphological classification illustrated as coloured points where the colour indicates the numeric morphological type T. The dashed line (slope 2.5) illustrates the line of constant disk luminosity, and the crosses at the top right illustrate the seven disky ellipticals from Scorza \& Bender (1995). At the top and right, the normalised distribution of each parameter is shown before (dotted) and after (solid) applying volume correction.} 
\label{fig:mu0rd}
\end{center}
\end{figure*}
\section{Freeman Law and \rd--$\mu_0$ Plane}
The Freeman Law defines an upper limit for $\mu_0$ and is hereby confirmed by our analysis of the largest sample ever studied in this context (see Fig.~\ref{fig:mu0rd}). However, disk galaxies with morphological type $T\ge6$ have fainter $\mu_0$. These results in $r$-band are comparable with those in other SDSS bands (Fathi 2010). Combined with our previous results, i.e. that \rd\ varies by two orders of magnitude independent of morphological type, this result implies that disks with large scale lengths, not necessarily have higher $\mu_0$.

The $\mu_0$ has a Gaussian distribution with $\langle\mu_0\rangle=20.2\pm0.7$ mag~arcsec$^{-2}$ with a linear trend seen in Fig.~\ref{fig:mu0rd} (applying different internal extinction parameters changes this mean value by 0.2 mag~arcsec$^{-2}$). The top right corner is enclosed by the constant disk luminosity line, void of objects. The top right corner is also the region where the disk luminosity exceeds $3L^\star$, thus the absence of galaxies in this region cannot be a selection effect since big bright galaxies cannot be missed in our diameter selected sample. However, it is clear that selection effect plays a role in populating the lower left corner of this diagram. Analogue to the \rd--$\mu_0$ plane, the Tully-Fisher relation implies lines of constant maximum speed a disk can reach. 

Our 282 well-classified galaxies (illustrated with coloured dots in Fig.~\ref{fig:mu0rd}) follow the results of, e.g., Graham and de Blok (2001), and confirm that disks of intermediate and early type spirals have higher $\mu_0$ while the late type spirals have lower $\mu_0$, and they populate the lower left corner of the diagram. Intermediate morphologies are mixed along a linear slope of 2.5 in the \rd--$\mu_0$ plane, coinciding with the region populated by S0s as shown by Kent (1985) and disky ellipticals shown by Scorza \& Bender (1995) The \rd, on the other hand, does not vary as a function of morphological type (Fathi et al. 2010). Investigating galaxy masses, we find a forth quantity is equally important in this analysis. The total galaxy mass separates the data along lines parallel to the dashed lines drawn in Fig.~\ref{fig:mu0rd}. This is indeed also confirmed by the Tully-Fisher relation. Moreover, we validate that the lower mass galaxies are those with type $\ge 6$. 

Investigation of the asymmetry and concentration in this context further confirms the expected trends, i.e. that these parameters increase for later types, and central stellar velocity dispersion decrease for later type spiral galaxies, however, we note that these correlations are well below one-sigma confidence level. However, the higher asymmetry galaxies populate a region more extended toward the bottom right corner with respect to the high asymmetry galaxies. The middle panel shows an opposite trend, and the bottom panel shows that larger velocity dispersion has the same effect as asymmetry (see Fathi 2010 for further details).

In the two relations analysed here, we find typically larger scatter than previous analyses, and although our sample represents bright disks, the sample size adds credibility to our findings. These results are fully consistent with the common understanding of the \rd--$\mu_0$ plane and the Freeman Law, and they contribute to past results since they are based on a sample which is two order of magnitudes greater than any previous study, with more than five times more late type spiral galaxies than any previous analysis.\\

{\em Acknowledgements: } I thank the IAU, LOC and my colleagues i the SOC for a stimulating symposium, and Mark Allen, Evanthia Hatziminaoglou, Thomas Boch, Reynier Peletier and Michael Gatchell for their invaluable input at various stages of this project.

\begin{discussion}

\discuss{R. Sanchez-Janssen}{How do you account for bulges or bars in your scale length determination?}

\discuss{K. Fathi}{We cut out the innermost 25\% of the expected disk size. This way we remove the contamination from the bulge light, and the bar we assume having a minor contribution to the disk light. The ultimate test to this simplified procedure is in the comparison between our derived scale lengths with those based on deep images and elaborate multi-component decomposition techniques, and we find a mean ratio (between our values and those from several authors) to be $\approx 1.0\pm0.2$. We see this as a good match, c.f., work by Knapen and van der Kruit in the early 90s.}

\end{discussion}


\begin{thebibliography}{}
\bibitem[Adelman-McCarthy et al. (2008)]{AMcetal08} Adelman-McCarthy, J. K. et al. 2008, ApJS, 175, 297
\bibitem[Disney (1976)]{Disney76} Disney, M. 1976, Nature, 263, 573
\bibitem[Graham \& de Blok (2001)]{GdeB01} Graham, A. W., de Blok, W. 2001, ApJ, 556, L177
\bibitem[Fathi et al. (2010)]{Fathietal10} Fathi, K. et al. 2010, MNRAS, 406, 1595
\bibitem[Fathi(2010)]{Fathi10} Fathi, K. 2010, ApJ, 722, L120
\bibitem[Freeman (1970)]{Freeman70} Freeman, K. C. 1970, 160, 811  
\bibitem[Kent (1985)]{Kent85} Kent, S. 1985, ApJSS, 59, 115 
\bibitem[Paturel et al. (2003)]{Petal03} Paturel G. et al. 2003, A\&A, 412, 45
\bibitem[Scorza \& Bender (1995)]{SB95} Scorza, C., Bender, R. 1995, A\&A, 293, 20 
\bibitem[White \& Rees (1978)]{WR78} White, S. D. M., Rees, M. J. 1978, MNRAS, 183, 341
\bibitem[York et al. (2000)]{Yorketal00} York, D. G. et al. 2000, AJ, 120, 1579
\end{thebibliography}
\end{document}